\documentclass[aps,pra,showpacs,superscriptaddress,preprint]{revtex4}
\usepackage{amsmath}
\usepackage{graphicx}

\catcode`ð=\active
 \defð{\u{g}}
 \catcode`Ð=\active
 \defÐ{\u{G}}
 \catcode`Ý=\active
\defÝ{\. I}
 \catcode`ö=\active
\defö{\"{o}}
 \catcode`Ö=\active
 \defÖ{\"O}
 \catcode`ü=\active
 \defü{\"{u}}
 \catcode`Ü=\active
 \defÜ{\"{U}}
 \catcode`Þ=\active
\defÞ{\c{S}}
 \catcode`þ=\active
 \defþ{\c{s}}
 \catcode`ý=\active
 \defý{{\i}}
 \catcode`ç=\active
\defç{\d{c}}
 \catcode`Ç=\active
\defÇ{\d{C}}

\input{tcilatex}

\begin{document}

\title{On the bound-state solutions of the Manning-Rosen potential including
improved approximation to the orbital centrifugal term}
\author{Sameer M. Ikhdair}
\email[E-mail: ]{sikhdair@neu.edu.tr}
\affiliation{Physics Department, Near East University, Nicosia, Mersin 10, Turkey}
\date{%
%TCIMACRO{\TeXButton{today}{\today}}%
%BeginExpansion
\today%
%EndExpansion
}

\begin{abstract}
The approximate analytical bound state solution of the Schr\"{o}dinger
equation for the Manning-Rosen potential is carried out by taking a new
approximation scheme to the orbital centrifugal term. The Nikiforov-Uvarov
method is used in the calculations. We obtain analytic forms for the energy
eigenvalues and the corresponding normalized wave functions in terms of the
Jacobi polynomials or hypergeometric functions for different screening
parameters $1/b$. The rotational-vibrational energy states for a few
diatomic molecules are calculated for arbitrary quantum numbers $n$ and $l$
with different values of the potential parameter $\alpha $. The present
numerical results agree within five decimal digits with the previously
reported results for different $1/b$ values. A few special cases of the $s$%
-wave ($l=0$) Manning-Rosen potential and the Hulth\'{e}n potential are also
studied.\newline
\end{abstract}

\maketitle

\newpage

\section{Introduction}

The exact analytic solutions of the wave equations (nonrelativistic and
relativistic) are only possible for certain potentials of physical interest
under consideration since they contain all the necessary information on the
quantum system. It is well known that the exact solutions of these wave
equations are only possible in a few simple cases such as the Coulomb, the
harmonic oscillator, the pseudoharmonic potentials and others [1-5]. The
analytic exact solutions of the wave equation with some exponential-type
potentials are impossible for $l\neq 0$ states. Therefore, approximation
schemes have to be used to deal with the orbital centrifugal term like the
Pekeris approximation [6-8] and the approximated scheme suggested by Greene
and Aldrich [9]. Some of these exponential-type potentials include the Morse
potential [10], the Hulth\'{e}n potential [11], the P\"{o}schl-Teller [12],
the Woods-Saxon potential [13], the Kratzer-type and pseudoharmonic
potentials [14], the Rosen-Morse-type potentials [15], the Manning-Rosen
potential [16-19] and other multiparameter exponential-type potentials
[20,21] etc.

The Manning-Rosen (MR) potential has been one of the most useful and
convenient models to study the energy eigenvalues of diatomic molecules
[16]. As an empirical potential, the MR potential gives an excellent
description of the interaction between the two atoms in a diatomic molecule
and also it is very reasonable in describing such interactions close to the
surface. The short range MR potential is defined by [16-19]

\begin{equation}
V(r)=\frac{\hbar ^{2}}{2\mu b^{2}}\left[ \frac{\alpha (\alpha -1)}{\left(
e^{r/b}-1\right) ^{2}}-\frac{A}{e^{r/b}-1}\right] ,
\end{equation}%
where $A$ and $\alpha $ are two constants and the parameter $b$
characterizes the range of the potential [22]. The above potential may be
further put in the following simple form

\begin{equation}
V(r)=-\frac{Ce^{r/b}+D}{\left( e^{r/b}-1\right) ^{2}},\text{ }C=A,\text{ }%
D=-A-\alpha \text{(}\alpha -1)\text{,}
\end{equation}%
which is usually used for the description of diatomic molecular vibrations
and rotations [23,24]. It is also used in several branches of physics for
their bound states and scattering properties. This potential remains
invariant by mapping $\alpha \rightarrow 1-\alpha $ and has a relative
minimum at $r_{0}=b\ln \left[ 1+2\alpha (\alpha -1)/A\right] $ with value $%
V(r_{0})=-\frac{\hbar ^{2}A^{2}}{8\mu b^{2}\alpha (\alpha -1)}$ for $\alpha
<0$ or $\alpha >1$ and $A>0.$ Moreover, the second derivative determines the
force constants at $r=r_{0}$ which is given by

\begin{equation}
\left. \frac{d^{2}V}{dr^{2}}\right\vert _{r=r_{0}}=\frac{A^{2}\left[
A+2\alpha (\alpha -1)\right] ^{2}}{8b^{4}\alpha ^{3}(\alpha -1)^{3}}.
\end{equation}%
If $\alpha =0$ or $\alpha =1,$ the potential (1) reduces to the Hulth\'{e}n
potential [11]. For the potential in Eq. (1) [16-19], the Schr\"{o}dinger
equation (\textrm{SE}) can be easily solved for the $s$-wave, angular
momentum quantum number $l=0.$ However, for the general solution, one needs
to include some approximations to obtain analytical or semi-analytical
solutions to the \textrm{SE}. Also, it is often necessary to determine the $%
l $-wave ($l\neq 0$ states), so an analytic procedure would be advantageous
[25-27]. Hence, in the previous papers, several approximations have been
developed to find better analytical formulas for the energy bound states and
wave functions. For instance, in the $l=0$ case, the bound-state energy
spectra for the MR potential have already been calculated by using the
path-integral approach [17] and function analysis method [18]. For the $%
l\neq 0$ case, the potential can not be solved exactly without using
approximation scheme. Recently, Qiang and Dong [19] approximated the
centrifugal term%
\begin{equation*}
\frac{1}{r^{2}}\approx \frac{1}{b^{2}}\left[ \frac{1}{e^{r/b}-1}+\frac{1}{%
\left( e^{r/b}-1\right) ^{2}}\right] =\frac{1}{b^{2}}\frac{e^{r/b}}{\left(
e^{r/b}-1\right) ^{2}}
\end{equation*}%
and studied $l$-wave bound-state solutions of the SE for MR potential.
Further, the scattering state solutions for the same potential and
approximation have also been investigated [25]. The above approximation has
also been applied to obtain the $l$-wave solutions of \textrm{SE} with the
MR potential in three-dimensions and $D$-dimensions and also with the Hulth%
\'{e}n potential using the Nikiforov and Uvarov (NU) method [11,19,26,27].
The present approximations provide good results which are in agreement
within five decimal digits with the previously reported numerical
integration method by Lucha and Sch\"{o}berl [28] for short-range potential
(large $b$ and small $l$) but not for long-range potential (small $b$ and
large $l$)$.$

The main purpose of the present paper is to improve the accuracy of the
previous approximations introduced in [26,29], so that we apply a different
approximation scheme recently proposed in Ref. [27] for the centrifugal term 
$l(l+1)r^{-2}$ to make the results in higher agreement with Ref. [28]. Thus,
with this new approximation scheme, we calculate the $l\neq 0$ energy levels
and wave functions for the MR potential using the NU method [30] which has
shown its power in calculating the exact energy levels for some solvable
quantum systems. For this, the results are in better agreement with those
obtained by means of numerical integration method [28]. As an illustration,
the method is applied to find the ro-vibrational energy states for a few
diatomic molecules: $\mathrm{HCl},$\textrm{\ }$\mathrm{CH},$\textrm{\ }$%
\mathrm{LiH},\mathrm{CO},$\textrm{\ }$\mathrm{NO},$\textrm{\ }$\mathrm{O}%
_{2},$\textrm{\ }$\mathrm{I}_{2},$\textrm{\ }$\mathrm{N}_{2},$\textrm{\ }$%
\mathrm{H}_{2}$ and $\mathrm{Ar}_{2}$.

The paper is organized as follows: In Section II, we apply the new
approximation scheme to calculate the $l$-wave bound state eigensolutions of
the \textrm{SE} for MR potential by using the NU method. In Section III, we
present our ro-vibrational energy levels for a few diatomic molecules.
Section IV, is devoted for two special cases, namely, $s$-wave ($l=0)$ and
the Hulth\'{e}n potential. Finally, we make a few concluding remarks in
Section V.

\section{Bound State Solutions}

To study any quantum physical system, we solve the original $\mathrm{SE}$
that is given in the well known textbooks [1,2] 
\begin{equation}
\left( \frac{p^{2}}{2m}+V(r)\right) \psi _{nlm}(\mathbf{r})=E_{nl}\psi
_{nlm}(\mathbf{r}),
\end{equation}%
where the potential $V(r)$ is taken as the MR form in (1). Further, we set
the wave functions $\psi _{nlm}(\mathbf{r})=\frac{u_{nl}(r)}{r}Y_{lm}(\theta
,\phi )$ to obtain the following radial Schr\"{o}dinger eqauation: 
\begin{subequations}
\begin{equation}
\frac{d^{2}u_{nl}(r)}{dr^{2}}+\left[ \frac{2\mu E_{nl}}{\hbar ^{2}}-V_{\text{%
eff}}(r)\right] u_{nl}(r)=0,
\end{equation}%
\begin{equation}
V_{\text{eff}}(r)=\frac{1}{b^{2}}\left[ \frac{\alpha (\alpha -1)}{\left(
e^{r/b}-1\right) ^{2}}-\frac{A}{e^{r/b}-1}\right] +\frac{l(l+1)}{r^{2}},
\end{equation}%
in which $u_{nl}(0)=0$ and $\underset{r\rightarrow \infty }{\lim }%
u_{nl}(r)=0.$ To solve the above equation for $l\neq 0$ states$,$ we need to
apply the following approximate scheme to the centrifugal term given by 
\end{subequations}
\begin{equation}
\frac{1}{r^{2}}\approx \frac{1}{b^{2}}\left[ D_{0}+D_{1}\frac{1}{e^{r/b}-1}%
+D_{2}\frac{1}{\left( e^{r/b}-1\right) ^{2}}\right] ,
\end{equation}%
and the higher order terms are neglected. These solutions are valid for $%
r/b\ll 1,$ that is, the solutions obtained are valid for $\alpha (\alpha
-1)/A\ll 1$ but positive. Obviously, the above approximation to the
centrifugal term turns to $r^{-2}$ when the parameter $b$ goes to infinity
(small screening parameter $\delta =1/b$) as%
\begin{equation}
\underset{b\rightarrow \infty }{\lim }\left[ \frac{1}{b^{2}}\left( D_{0}+%
\frac{1}{e^{r/b}-1}+\frac{1}{\left( e^{r/b}-1\right) ^{2}}\right) \right] =%
\frac{1}{r^{2}},  \tag{6a}
\end{equation}%
which shows that the usual approximation is the limit of our approximation
(cf. e.g., [31] and the references therein). The values of the parameters $%
D_{i}$ $(i=0,1$ and $2)$ are given by [27,31]%
\begin{equation}
D_{0}\simeq \frac{1}{12},\text{ }D_{1}=\text{ }D_{2}=1.
\end{equation}%
However, the values of the parameters $D_{i}$ $(i=0,1$ and $2)$ used by Wei
and Dong \ [32] are given by 
\begin{subequations}
\begin{equation}
D_{0}=\frac{12\epsilon _{1}^{2}-4\epsilon _{1}\left( 2A+3\epsilon
_{1}\right) \log (\epsilon _{2})+\epsilon _{3}^{2}\log (\epsilon _{2})^{2}}{%
\epsilon _{4}^{2}\log (\epsilon _{2})^{4}},
\end{equation}%
\begin{equation}
D_{1}=\frac{8\epsilon _{1}^{2}\left[ -6\epsilon _{1}+\left( 3A+4\epsilon
_{1}\right) \log (\epsilon _{2})\right] }{A\epsilon _{4}^{2}\log (\epsilon
_{2})^{4}},
\end{equation}%
\begin{equation}
D_{2}=-\frac{16\epsilon _{1}^{3}\left[ -3\epsilon _{1}+\epsilon _{3}\log
(\epsilon _{2})\right] }{A^{2}\epsilon _{4}^{2}\log (\epsilon _{2})^{4}},
\end{equation}%
where $\epsilon _{1}=\alpha (\alpha -1),$ $\epsilon _{2}=1+2\alpha (\alpha
-1)/A,$ $\epsilon _{3}=A\epsilon _{2}$ and $\epsilon _{4}=b\epsilon _{3}.$

Now, we need to recast differential equation (5) and the approximation (6)
into the form of Eq. (1) of Ref. [33] by introducing the change in the
variables $r\rightarrow z$ through the mapping function $z=e^{-r/b},$ and
defining 
\end{subequations}
\begin{subequations}
\begin{equation}
\text{ }\varepsilon _{nl}=\sqrt{-\frac{2\mu b^{2}E_{nl}}{\hbar ^{2}}+\Delta
E_{l}}>0,\text{ }E_{nl}<\frac{\hbar ^{2}}{2\mu b^{2}}\Delta E_{l},\text{ }%
\Delta E_{l}=l(l+1)D_{0},
\end{equation}%
\begin{equation}
\beta _{1}=A-l(l+1)D_{1},
\end{equation}%
\begin{equation}
\beta _{2}=\alpha (\alpha -1)+l(l+1)D_{2},
\end{equation}%
in order to obtain the following compact hypergeometric equation: 
\end{subequations}
\begin{equation*}
\frac{d^{2}u_{nl}(z)}{dz^{2}}+\frac{(1-z)}{z(1-z)}\frac{du_{nl}(z)}{dz}
\end{equation*}%
\begin{equation}
+\frac{1}{\left[ z(1-z)\right] ^{2}}\left\{ -\varepsilon _{nl}^{2}{}+\left(
2\varepsilon _{nl}^{2}+\beta _{1}\right) z-\left( \varepsilon
_{nl}^{2}+\beta _{1}+\beta _{2}\right) z^{2}\right\} u_{nl}(z)=0.
\end{equation}%
We notice that for the presence of bound state (real) solutions, $%
\varepsilon _{nl}$ must be a positive real parameter and we require that%
\begin{equation}
z=\left\{ 
\begin{array}{ccc}
0, & \text{when} & r\rightarrow \infty , \\ 
1, & \text{when} & r\rightarrow 0,%
\end{array}%
\right.
\end{equation}%
for the radial wave functions to fulfill the boundary conditions, i.e., $%
u_{nl}(0)\rightarrow 0$ and $u_{nl}(1)\rightarrow 0.$ Let us begin by
comparing Eq. (10) with Eq. (1) of Ref. [33], then we obtain the following
definitions: 
\begin{equation}
\widetilde{\tau }(z)=1-z,\text{\ }\sigma (z)=z-z^{2},\text{\ }\widetilde{%
\sigma }(z)=-\varepsilon _{nl}^{2}{}+\left( 2\varepsilon _{nl}^{2}+\beta
_{1}\right) z-\left( \varepsilon _{nl}^{2}+\beta _{1}+\beta _{2}\right)
z^{2}.
\end{equation}%
After applying the relations (A1-A4) of Ref. [33], the following useful
functions usually defined by the NU method [30] are achieved%
\begin{equation}
k=\beta _{1}-a\varepsilon _{nl},\text{ \ }a=\sqrt{(1-2\alpha
)^{2}+4l(l+1)D_{2}}.
\end{equation}%
\begin{equation}
\text{\ }\pi (z)=-\frac{z}{2}-\frac{1}{2}\left[ \left( a+2\varepsilon
_{nl}\right) z-a\right] ,
\end{equation}%
and%
\begin{equation}
\tau (z)=1+2\varepsilon _{nl}-\left( 2+2\varepsilon _{nl}+a\right) z,\text{ }%
\tau ^{\prime }(z)=-\left( 2+2\varepsilon _{nl}+a\right) .
\end{equation}%
We can also write the values of $\lambdabar =k+\pi ^{\prime }(z)$ and $%
\lambdabar _{n}=-n\tau ^{\prime }(z)-\frac{n\left( n-1\right) }{2}\sigma
^{\prime \prime }(z),$\ $n=0,1,2,\cdots $ to obtain%
\begin{equation}
\lambdabar =\beta _{1}-(1+a)\left( \frac{1}{2}+\varepsilon _{nl}\right) ,
\end{equation}%
and%
\begin{equation}
\lambdabar _{n}=n(1+n+a+2\varepsilon _{nl}),\text{ }n=0,1,2,\cdots ,
\end{equation}%
respectively. Furthermore, using the relation, $\lambdabar =\lambdabar _{n},$
or alternatively the energy equation (A5) of Ref. [33]$,$ allows one to
obtain%
\begin{equation}
\varepsilon _{nl}=\frac{A+\alpha (\alpha -1)+l(l+1)\left( D_{2}-D_{1}\right) 
}{2n+1+a}-\frac{2n+1+a}{4}.
\end{equation}%
Plugging the parameters given in Eq. (9) into Eq. (18), we finally obtain
the following discrete bound-state energy eigenvalues:%
\begin{equation*}
E_{nl}=E_{nl}^{\mathrm{(approx)}}=\frac{\hbar ^{2}l(l+1)D_{0}}{2\mu b^{2}}%
\text{ }
\end{equation*}%
\begin{equation}
-\frac{\hbar ^{2}}{2\mu b^{2}}\left[ \frac{A+\alpha (\alpha -1)+l(l+1)\left(
D_{2}-D_{1}\right) }{2n+1+\sqrt{(1-2\alpha )^{2}+4l(l+1)D_{2}}}-\frac{2n+1+%
\sqrt{(1-2\alpha )^{2}+4l(l+1)D_{2}}}{4}\right] ^{2},
\end{equation}%
where $0\leq n\leq n_{\max }$ and $l=0,1,2,\cdots ,$ signify the usual
vibrational and rotational quantum numbers, respectively. It is found that
the parameter $a$ in Eq. (13) remains invariant by mapping $\alpha
\rightarrow 1-\alpha ,$ so do the bound state energies $E_{nl}.$ An
important quantity of interest for the MR potential is the critical coupling
constant $A_{c},$ which is that value of $A$ for which the binding energy of
the level in question becomes zero. Hence, using Eq. (19), in atomic units $%
\hbar ^{2}=\mu =Z=e=1,$ we find the following critical coupling constant%
\begin{equation}
A_{c}=\frac{1}{4}\left( 2n+1+\sqrt{(1-2\alpha )^{2}+4l(l+1)D_{2}}\right)
^{2}-\alpha (\alpha -1)-l(l+1)\left( D_{2}-D_{1}\right) .
\end{equation}%
Let us now turn to the calculations of the radial part of the normalized
wave functions. After applying the relations (A6-A10) of Ref. [33], we obtain%
\begin{equation}
\phi (z)=z^{\varepsilon _{nl}}(1-z)^{\left( 1+a\right) /2},
\end{equation}%
\begin{equation}
\rho (z)=z^{2\varepsilon _{nl}}(1-z)^{a},
\end{equation}%
\begin{equation}
y_{nl}(z)=C_{n}z^{-2\varepsilon _{nl}}(1-z)^{-a}\frac{d^{n}}{dz^{n}}\left[
z^{n+2\varepsilon _{nl}}(1-z)^{n+a}\right] .
\end{equation}%
The functions $\ y_{nl}(z)$, up to a numerical factor, are in the form of\
Jacobi polynomials, i.e., $\ y_{nl}(z)\simeq P_{n}^{(2\varepsilon
_{nl},a)}(1-2z)$ (the physical interval $(0,\infty )$ for variable $r$ is
mapped to the interval ($0,1)$ for variable $z)$ [13,14]. Hence, the
approximated radial wave functions satisfying Eq. (5) are given by%
\begin{equation*}
u_{nl}(r)=u_{nl}^{\mathrm{(approx)}}(r)=\mathcal{N}_{nl}e^{-\varepsilon
_{nl}r/b}(1-e^{-r/b})^{\nu _{l}}%
\begin{array}{c}
_{2}F_{1}%
\end{array}%
\left( -n,n+2\left( \varepsilon _{nl}+\nu _{l}\right) ;2\varepsilon
_{nl}+1;e^{-r/b}\right) ,
\end{equation*}%
\begin{equation}
\varepsilon _{nl}>0,\text{ }\nu _{l}=\left( 1+a\right) /2\geq 1,
\end{equation}%
where $a$ and $\varepsilon _{nl}$ are given in Eqs. (13) and (18),
respectively and $\mathcal{N}_{nl}$ is a normalization constant determined
in the Appendix B.

When $l=0,$ we deal with $s$-wave case, the possible energies for the bound
states and the corresponding wave functions are written explicitly, for $%
\alpha <1/2$:%
\begin{equation}
\left\{ 
\begin{array}{c}
E_{n}=-\frac{\hbar ^{2}}{8\mu b^{2}}\left[ \frac{A+\alpha (\alpha -1)}{%
n-\alpha +1}-\left( n-\alpha +1\right) \right] ^{2}; \\ 
n=0,1,2,\cdots ,n_{\max }=\left[ \sqrt{A+\alpha (\alpha -1)}+\alpha -1\right]
\text{ and} \\ 
u_{n0}(r)=\mathcal{N}_{n}e^{-\left( \varepsilon _{n}/b\right)
r}(1-e^{-r/b})^{\left( 1-\alpha \right) } \\ 
\begin{array}{c}
_{2}F_{1}%
\end{array}%
\left( -n,n+2\left( \varepsilon _{n}-\alpha +1\right) ;2\varepsilon
_{n}+1;e^{-r/b}\right) ,%
\end{array}%
\right. \text{ }
\end{equation}%
where $\varepsilon _{n}=\frac{1}{2}\left[ \frac{A+\alpha (\alpha -1)}{%
n-\alpha +1}-\left( n-\alpha +1\right) \right] $ and for $\alpha >1/2$:%
\begin{equation}
\left\{ 
\begin{array}{c}
E_{n}=-\frac{\hbar ^{2}}{8\mu b^{2}}\left[ \frac{A+\alpha (\alpha -1)}{%
n+\alpha }-\left( n+\alpha \right) \right] ^{2}; \\ 
n=0,1,2,\cdots ,\text{ }n_{\max }=\left[ \sqrt{A+\alpha (\alpha -1)}-\alpha %
\right] \text{ and} \\ 
u_{n0}(r)=\widetilde{\mathcal{N}}_{n}e^{-\left( \varepsilon _{n}^{\prime
}/b\right) r}(1-e^{-r/b})^{\alpha } \\ 
\begin{array}{c}
_{2}F_{1}%
\end{array}%
\left( -n,n+2\left( \varepsilon _{n}^{\prime }+\alpha \right) ;2\varepsilon
_{n}^{\prime }+1;e^{-r/b}\right) ,%
\end{array}%
\right. \text{ }
\end{equation}%
where $\varepsilon _{n}^{\prime }=\frac{1}{2}\left[ \frac{A+\alpha (\alpha
-1)}{n+\alpha }-\left( n+\alpha \right) \right] .$ The normalization
constants $\mathcal{N}_{n}$ and $\widetilde{\mathcal{N}}_{n}$ are calculated
explicitly in the Appendix B. Notice that $n_{\max }$ is the number of bound
states for the whole bound spectrum near the continuous zone. $n_{\max }$ is
the largest integer which is less than or equal to the value of $n$ that
makes the right side of Eqs. (25) and (26) to vanish. The above results are
in identical to Eqs. (12) and (13) given by Ref. [34].

\section{Applications to Diatomic Molecules}

To show the accuracy of the new approximation scheme, we have calculated the
ro-vibrational energy spectra for various $n$ and $l$ quantum numbers with
two different values of the parameters $\alpha .$ The results obtained by
means of Eq. (19) are compared with those obtained by a MATHEMATICA package
programmed by Lucha and Sch\"{o}berl [28] as listed in Table 1 for
short-range (large $b$) and long-range (small $b$) potentials. This is an
illustration to assess the validity and usefulness of our present
approximations. The results of the energy spectrum for $p$-state show that
the percentage accuracy decreases as either $n$ or $1/b$ increases, for
example, when $1/b=0.025,$ then the range of accuracies can be as follows: $%
0.00075\%,$ $0.00087\%,$ $0.0014\%,$ $0.017\%$ and $0.11\%$ for $n=0,1,2,3,$
and $4,$ respectively. However, when $1/b$ changes between $0.025-0.075,$
then the range of accuracies can be $0.00075\%-0.0022\%,$ $%
0.00087\%-0.068\%, $ $0.0014\%-1.57\%$ for $n=0,1$ and $2,$ respectively. In
addition, we present the ro-vibrational energy states for a few diatomic
molecules $\mathrm{HCl},$ $\mathrm{CH},$ $\mathrm{LiH},$ $\mathrm{CO},$ $%
\mathrm{NO},$ $\mathrm{O}_{2},$ $\mathrm{I}_{2},$ $\mathrm{N}_{2},$ $\mathrm{%
H}_{2}$ and $\mathrm{Ar}_{2}$ in Tables 2-6. Lowest eigenvalues of $%
l=0,1,2,3 $ are given at four values of $1/b$ in the range $0.025-0.1$
covering both weaker and stronger interaction to demonstrate the generality
of our results. The formalism is quite simple, computationally efficient,
reliable and illustrated very accurate.

\section{Some Special Cases}

Let us study a few special cases. In the case where $\alpha =0$ or $\alpha
=1,$ the MR potential $(1)$ reduces to the Hulth\'{e}n potential [9,11]:%
\begin{equation}
V^{(H)}(r)=-V_{0}\frac{e^{-\delta r}}{1-e^{-\delta r}},\text{ }%
V_{0}=Ze^{2}\delta ,\text{ }\delta =b^{-1},
\end{equation}%
where $Ze^{2}$ is the strength and $\delta $ is the screening parameter and $%
b$ is the range of potential. If the potential is used for atoms, the $Z$ is
identified with the atomic number. Furthermore, if taking $b=1/\delta $ and
identifying $\left( A\hbar ^{2}/2\mu b^{2}\right) $ as $Ze^{2}\delta ,$ we
are able to obtain the ro-vibrating energy states and the normalized
wavefunctions deduced from Eqs. (19) and (24), respectively,%
\begin{equation*}
E_{nl}=-\frac{\hbar ^{2}\delta ^{2}}{2\mu }\left[ \frac{\left( 2\mu
Ze^{2}/\hbar ^{2}\delta \right) +l(l+1)\left( D_{2}-D_{1}\right) }{2n+1+%
\sqrt{1+4l(l+1)D_{2}}}-\frac{2n+1+\sqrt{1+4l(l+1)D_{2}}}{4}\right] ^{2}
\end{equation*}%
\begin{equation}
+\frac{\hbar ^{2}\delta ^{2}l(l+1)D_{0}}{2\mu },\text{ }0\leq n,l<\infty ,
\end{equation}%
and%
\begin{equation*}
u_{nl}(r)=\mathcal{N}_{nl}e^{-\left( \varepsilon _{nl}/b\right)
r}(1-e^{-r/b})^{\nu _{l}}%
\begin{array}{c}
_{2}F_{1}%
\end{array}%
\left( -n,n+2\left( \varepsilon _{nl}+\nu _{l}\right) ;2\varepsilon
_{nl}+1;e^{-r/b}\right) ,
\end{equation*}%
\begin{equation}
\varepsilon _{nl}=\sqrt{-\frac{2\mu E_{n,l}}{\hbar ^{2}\delta ^{2}}%
+l(l+1)D_{0}}>0,\text{ }\nu _{l}=\frac{1}{2}\left( 1+\sqrt{1+4l(l+1)D_{2}}%
\right) \geq 1,
\end{equation}%
where $\mathcal{N}_{nl}$ is given in the Appendix B. Also, for $s$-wave ($%
l=0)$ states, we get%
\begin{equation}
E_{n}=-\frac{\mu \left( Ze^{2}\right) ^{2}}{2\hbar ^{2}}\left[ \frac{1}{(n+1)%
}-\frac{\hbar ^{2}\delta }{2Ze^{2}\mu }(n+1)\right] ^{2},\text{ }0\leq
n<\infty .
\end{equation}%
and%
\begin{equation*}
u_{n}(r)=\mathcal{N}_{n}e^{-\left( \varepsilon _{n}/b\right) r}(1-e^{-r/b})%
\begin{array}{c}
_{2}F_{1}%
\end{array}%
\left( -n,n+2\left( \varepsilon _{n}+1\right) ;2\varepsilon
_{n}+1;e^{-r/b}\right) ,
\end{equation*}%
\begin{equation}
\varepsilon _{n}=\sqrt{-\frac{2\mu E_{n0}}{\hbar ^{2}\delta ^{2}}}>0,
\end{equation}%
where $\mathcal{N}_{n}$ can be easily found from either relation (B7) or
(B9) after setting $\alpha =0$ or $\alpha =1$ in the Appendix B$,$
respectively. Here in this case $\varepsilon _{n}=\varepsilon _{n}^{\prime }$
and the number of bound states $n_{\max }$ is also same in both relations
(B8) and (B10). In the usual approximation [19] where $D_{0}=0$ and $%
D_{1}=D_{2}=1,$ Eqs. (28) and (29) turn out to become%
\begin{equation}
E_{nl}=-\frac{\mu \left( Ze^{2}\right) ^{2}}{2\hbar ^{2}}\left[ \frac{1}{%
\left( n+l+1\right) }-\frac{\hbar ^{2}\delta }{2\mu Ze^{2}}\left(
n+l+1\right) \right] ^{2},\text{ }0\leq n,l<\infty ,
\end{equation}%
and%
\begin{equation*}
u_{nl}(r)=\mathcal{N}_{nl}e^{-\left( \varepsilon _{nl}/b\right)
r}(1-e^{-r/b})^{l+1}%
\begin{array}{c}
_{2}F_{1}%
\end{array}%
\left( -n,n+2\left( \varepsilon _{nl}+l+1\right) ;2\varepsilon
_{nl}+1;e^{-r/b}\right) ,
\end{equation*}%
\begin{equation}
\varepsilon _{nl}=\sqrt{-\frac{2\mu E_{nl}}{\hbar ^{2}\delta ^{2}}}>0,
\end{equation}%
where $\mathcal{N}_{nl}$ can be found from relation (B6) by setting $\nu
_{l}=l+1.$ Essentially, these results coincide with those obtained by the
Feynman integral method [17] and the standard way [18,19]. In following Ref.
[27] by taking $D_{1}=D_{2}=1$ and $D_{0}=1/12,$ Eqs. (28) and (29) turn out
to become 
\begin{equation}
E_{nl}=-\frac{\mu \left( Ze^{2}\right) ^{2}}{2\hbar ^{2}}\left[ \frac{1}{%
\left( n+l+1\right) }-\frac{\hbar ^{2}\delta }{2\mu Ze^{2}}\left(
n+l+1\right) \right] ^{2}+\frac{l(l+1)\hbar ^{2}\delta ^{2}}{24\mu },\text{ }%
0\leq n,l<\infty ,
\end{equation}%
and%
\begin{equation*}
u_{nl}(r)=\mathcal{N}_{nl}e^{-\left( \varepsilon _{nl}/b\right)
r}(1-e^{-r/b})^{l+1}%
\begin{array}{c}
_{2}F_{1}%
\end{array}%
\left( -n,n+2\left( \varepsilon _{nl}+l+1\right) ;2\varepsilon
_{nl}+1;e^{-r/b}\right) ,
\end{equation*}%
\begin{equation}
\varepsilon _{nl}=\sqrt{-\frac{2\mu E_{nl}}{\hbar ^{2}\delta ^{2}}+\frac{%
l(l+1)}{12}}>0,
\end{equation}%
which coincide for the ground state with G\"{o}n\"{u}l \textit{et al.} [9]
in Eq. (6). The Hulth\'{e}n potential behaves like the Coulomb potential
near the origin $(r\rightarrow 0),$ but in the asymptotic region $(r\gg 1)$
the Hulth\'{e}n potential decreases exponentially, so its capacity for bound
states is smaller than the Coulomb potential. However, for small values of
the screening parameter or for $\delta r\ll 1$ (i.e., $r/b\ll 1),$ the Hulth%
\'{e}n potential becomes the Coulomb potential given by : $V_{C}(r)=-\frac{%
Ze^{2}}{r}$ with energy levels and wave functions:

\begin{equation*}
E_{nl}=-\frac{\varepsilon _{0}}{(n+l+1)^{2}},\text{ }n,l=0,1,2,...
\end{equation*}

\begin{equation}
\varepsilon _{0}=\frac{Z^{2}\hbar ^{2}}{2\mu a_{0}^{2}},\text{ }a_{0}=\frac{%
\hbar ^{2}}{\mu e^{2}}
\end{equation}%
where $\varepsilon _{0}=13.6$ $eV$ and $a_{0}$ is Bohr radius for the
Hydrogen atom [3]. The wave functions also take the form: 
\begin{equation}
u_{nl}(r)=N_{nl}\exp \left[ -\frac{\mu Ze^{2}}{\hbar ^{2}}\frac{r}{\left(
n+l+1\right) }\right] r^{l+1}P_{n}^{\left( \frac{2\mu Ze^{2}}{\hbar
^{2}\delta (n+l+1)},2l+1\right) }(r),
\end{equation}%
which are found identical to Refs. [11,13].

\section{Coclusions}

We have applied an alternative improved approximation scheme of the
centrifugal potential $l(l+1)r^{-2}$ to obtain the energy levels and
corresponding wavefunctions for the MR potential in the framework of the NU
method for arbitrary $l$-waves. We have calculated the bound state energy
eigenvalues for the MR potential with $\alpha =0.75,1.5$ and $A=2b$ and
several $1/b$ screening paramete values. The wave functions are physical and
bound state energies are in good agreement with the results obtained by
other methods for short-range potential, small $\alpha $ and $l.$ The
precision of the resulting approximation of the wave functions (24) for the $%
V_{\text{eff}}(r)$ in Eq. (5b) is due to approximative character of the
centrifugal term $1/r^{2}$ in Eq. (6) for $l\neq 0$ states since the wave
functions are relevant to the bound state energy approximation in Eq. (19).
The approximation (6) for the centrifugal potential allows to get analytic
approximation (34) and (35) for the eigenvalues and the eigenfunctions for
the MR potential in the framework of the NU method for arbitrary $l$-waves.
It is not possible to compute the residual (the error in the solution $%
u_{nl}^{\mathrm{(approx)}}(r)$ given by Eq. (24)) since the correct (exact)
wave functions, $u_{nl}^{\mathrm{(correct)}}(r)$ of Eq. (5) are still not
found. Hence, the notation residual can be used for $R=Hu_{nl}^{\mathrm{%
(approx)}}(r)-E_{nl}^{\mathrm{(approx)}}u_{nl}^{\mathrm{(approx)}}(r)$ and
the error (or deviation) for the difference $u_{nl}^{\mathrm{(exact)}%
}(r)-u_{nl}^{\mathrm{(approx)}}(r)$ and $E_{nl}^{\mathrm{(exact)}}-E_{nl}^{%
\mathrm{(approx)}}.$ Due to the slowness of the numerical calculation of the
Hypergeometric functions $%
\begin{array}{c}
_{2}F_{1}%
\end{array}%
\left( -n,n+2\left( \varepsilon _{nl}+\nu _{l}\right) ;2\varepsilon
_{nl}+1;e^{-r/b}\right) $ and their derivatives in MATHEMATICA, the residual 
$R$ is not evaluated. This residual is expected to be $6$ order of magnitude
smaller than typical values $E_{nl}^{\mathrm{(approx)}}u_{nl}^{\mathrm{%
(approx)}}(r).$ Accordingly, the error $u_{nl}^{\mathrm{(exact)}}(r)-u_{nl}^{%
\mathrm{(approx)}}(r)$ is expected to be also small. Furthermore, the error
of approximation of the Hamiltonian (4) with potential (5) is already
smaller, since the approximation used in (6) is only valid when $r\ll b$
(small screening parameter $\delta =1/b$). In order to demonstrate this, NU
results have been compared with the results of the numerical integration
procedures using the MATHEMATICA program [28] and the results obtained from
usual approximations scheme of the centrifugal potential [26]. For small $%
1/b $ values$,$ NU results are in high agreement with the ones obtained in
[28], but in the high screening region (large $1/b$ values) the agreement is
poor. It is obvious from Table 1 that five (three) decimal digits are
expected to be correct in the present (previous) approximation. The reason
is simply that when $r/b$ increases in the high-screening region, the
agreement between the approximation expression and the centrifugal potential
decreases. We have also studied two special cases for $l=0,$ $l\neq 0$ and
Hulth\'{e}n potential. As we have seen, NU method puts no constraint on the
potential parameter values involved and is easy to implement. Our results
are sufficiently accurate for the practical purposes. Therefore, we have
applied the present solution in Eq. (19) to obtain the ro-vibrational
energies ($-E_{nl}$) for the $\mathrm{HCl},$ $\mathrm{CH},$ $\mathrm{LiH},%
\mathrm{CO},$ $\mathrm{NO},$ $\mathrm{O}_{2},$ $\mathrm{I}_{2},$ $\mathrm{N}%
_{2},$ $\mathrm{H}_{2}$ and $\mathrm{Ar}_{2}$ diatomic molecules.

\acknowledgments The author wishes to thank the anonymous referees for their
enlightening comments and suggestions. The support provided by the
Scientific and Technological Research Council of Turkey (T\"{U}B\.{I}TAK) is
highly appreciated.\bigskip

\appendix

\section{Normalization for the radial wave functions}

The normalization constant, $\mathcal{N}_{nl}$ can be determined in closed
form. We start by using the relation between the hypergeometric function and
the Jacobi polynomials (see formula (8.962.1) in [35]):%
\begin{equation*}
\begin{array}{c}
_{2}F_{1}%
\end{array}%
\left( -n,n+\nu +\mu +1;\nu +1;\frac{1-x}{2}\right) =\frac{n!}{\left( \nu
+1\right) _{n}}P_{n}^{\left( \nu ,\mu \right) }(x),
\end{equation*}%
\begin{equation}
\left( \nu +1\right) _{n}=\frac{\Gamma (n+\nu +1)}{\Gamma (\nu +1)},
\end{equation}%
to rewrite the wave functions in (24) as%
\begin{equation}
u_{nl}(r)=\mathcal{N}_{nl}\frac{n!\Gamma (2\varepsilon _{nl}+1)}{\Gamma
(n+2\varepsilon _{nl}+1)}e^{-\varepsilon _{nl}r/b}(1-e^{-r/b})^{\nu
_{l}}P_{n}^{(2\varepsilon _{nl},2\nu _{l}-1)}(1-2e^{-r/b}).
\end{equation}%
From the normalization condition $\int_{0}^{\infty }\left[ u_{nl}(r)\right]
^{2}dr=1$ and under the coordinate change $x=1-2e^{-r/b},$ the normalization
constant in (B2) is given by%
\begin{equation}
\mathcal{N}_{nl}^{-2}=b\left[ \frac{n!\Gamma (2\varepsilon _{nl}+1)}{\Gamma
(n+2\varepsilon _{nl}+1)}\right] ^{2}\int_{-1}^{1}\left( \frac{1-x}{2}%
\right) ^{2\varepsilon _{nl}}\left( \frac{1+x}{2}\right) ^{2\nu
_{l}-1}\left( \frac{1+x}{2}\right) \left[ P_{n}^{(2\varepsilon _{nl},2\nu
_{l}-1)}(x)\right] ^{2}dx.
\end{equation}%
The calculation of this integral can be done by writting 
\begin{equation*}
\frac{1+x}{2}=1-\left( \frac{1-x}{2}\right) ,
\end{equation*}%
and by making use of the following two integrals (see formula (7.391.5) in
[35]):%
\begin{equation}
\int_{-1}^{1}\left( 1-x\right) ^{\nu -1}\left( 1+x\right) ^{\mu }\left[
P_{n}^{\left( \nu ,\mu \right) }(x)\right] ^{2}dx=2^{\nu +\mu }\frac{\Gamma
(n+\nu +1)\Gamma (n+\mu +1)}{n!\nu \Gamma (n+\nu +\mu +1)},
\end{equation}%
which is valid for $\mathcal{R}$e($\nu )>0$ and $\mathcal{R}$e($\mu )>-1$
and (see formula (7.391.1) in [35]):%
\begin{equation}
\int_{-1}^{1}\left( 1-x\right) ^{\nu }\left( 1+x\right) ^{\mu }\left[
P_{n}^{\left( \nu ,\mu \right) }(x)\right] ^{2}dx=2^{\nu +\mu +1}\frac{%
\Gamma (n+\nu +1)\Gamma (n+\mu +1)}{n!\Gamma (n+\nu +\mu +1)(2n+\nu +\mu +1)}%
,
\end{equation}%
which is valid for $\mathcal{R}$e($\nu )>-1,$ $\mathcal{R}$e($\mu )>-1.$
This leads to%
\begin{equation}
\mathcal{N}_{nl}=\frac{1}{\Gamma (2\varepsilon _{nl}+1)}\left[ \frac{%
\varepsilon _{nl}(n+\varepsilon _{nl}+\nu _{l})}{2b(n+\nu _{l})}\frac{\Gamma
(n+2\varepsilon _{nl}+1)\Gamma (n+2\varepsilon _{nl}+2\nu _{l})}{n!\Gamma
\left( n+2\nu _{l}\right) }\right] ^{1/2},
\end{equation}%
where $0\leq n,l<\infty .$ In the $s$-wave $\left( l=0\right) $ case, the
above result is written explicitly, for $\alpha <1/2$: 
\begin{equation}
\mathcal{N}_{n}=\frac{1}{\Gamma (2\varepsilon _{n}+1)}\left[ \frac{%
\varepsilon _{n}(n+\varepsilon _{n}-\alpha +1)}{2b(n-\alpha +1)}\frac{\Gamma
(n+2\varepsilon _{n}+1)\Gamma (n+2\varepsilon _{n}-2\alpha +2)}{n!\Gamma
\left( n-2\alpha +2\right) }\right] ^{1/2},
\end{equation}%
where%
\begin{equation}
\varepsilon _{n}=\frac{A+\alpha (\alpha -1)}{2\left( n-\alpha +1\right) }-%
\frac{n-\alpha +1}{2},\text{ }0\leq n<n_{\max }=\left[ \sqrt{A+\alpha
(\alpha -1)}+\alpha -1\right]
\end{equation}%
in which $\alpha =0$ is included in $\left( -\infty ,1/2\right) $ and for $%
\alpha >1/2$:%
\begin{equation}
\widetilde{\mathcal{N}}_{n}=\frac{1}{\Gamma (2\varepsilon _{n}+1)}\left[ 
\frac{\varepsilon _{n}^{\prime }(n+\varepsilon _{n}^{\prime }+\alpha )}{%
2b(n+\alpha )}\frac{\Gamma (n+2\varepsilon _{n}^{\prime }+1)\Gamma
(n+2\varepsilon _{n}^{\prime }+2\alpha )}{n!\Gamma \left( n+2\alpha \right) }%
\right] ^{1/2},
\end{equation}%
where%
\begin{equation}
\varepsilon _{n}^{\prime }=\frac{A+\alpha (\alpha -1)}{2\left( n+\alpha
\right) }-\frac{n+\alpha }{2},\text{ }0\leq n<n_{\max }=\left[ \sqrt{%
A+\alpha (\alpha -1)}-\alpha \right] .
\end{equation}%
in which $\alpha =1$ is included in $\left( 1/2,\infty \right) .$

\newpage

{\normalsize %center
}

\newpage \baselineskip= 2\baselineskip% double space the text
%\end{document}
\bigskip \newpage

\begin{table}[tbp]
\caption{Bound state energy spectrum ($-E_{nl}$) (in atomic units) for the
Manning-Rosen potential as a function of $1/b$ for $%
2p,3p,3d,4p,4d,4f,5p,5d,5f,5g,6p,6d,6f$ and $6g$ states with $\protect\alpha %
=0.75,$ $\protect\alpha =1.5$ and $A=2b.$}%
\begin{tabular}{llllllll}
\tableline &  & $\alpha =0.75$ &  &  & $\alpha =1.5$ &  &  \\ 
states & $1/b$ & present & previous [26] & Lucha et al [28] & present & 
previous [26] & Lucha et al [28] \\ 
\tableline$2p$ & $0.025$ & $0.1205279$ & $0.1205793$ & $0.1205271$ & $%
0.0899715$ & $0.0900229$ & $0.0899708$ \\ 
& $0.050$ & $0.1082170$ & $0.1084228$ & $0.1082151$ & $0.0800414$ & $%
0.0802472$ & $0.0800400$ \\ 
& $0.075$ & $0.0964490$ & $0.0969120$ & $0.0964469$ & $0.0705703$ & $%
0.0710332$ & $0.0705701$ \\ 
& $0.100$ & $0.0852240$ & $0.0860740$ &  & $0.0615579$ & $0.0577157$ &  \\ 
$3p$ & $0.025$ & $0.0458783$ & $0.0459297$ & $0.0458779$ & $0.0369137$ & $%
0.0369651$ & $0.0369134$ \\ 
& $0.050$ & $0.0350614$ & $0.0352672$ & $0.0350633$ & $0.0272662$ & $%
0.0274719$ & $0.0272696$ \\ 
& $0.075$ & $0.0255480$ & $0.0260110$ & $0.0255654$ & $0.0189220$ & $%
0.0193850$ & $0.0189474$ \\ 
& $0.100$ & $0.0173379$ & $0.0181609$ &  & $0.0118813$ & $0.0127043$ &  \\ 
$3d$ & $0.025$ & $0.0447756$ & $0.0449299$ & $0.0447743$ & $0.0394801$ & $%
0.0396345$ & $0.0394789$ \\ 
& $0.050$ & $0.0336909$ & $0.0343082$ & $0.0336930$ & $0.0294456$ & $%
0.0300629$ & $0.0294496$ \\ 
& $0.075$ & $0.0237279$ & $0.0251168$ & $0.0237621$ & $0.0204232$ & $%
0.0218121$ & $0.0204663$ \\ 
$4p$ & $0.025$ & $0.0208094$ & $0.0208608$ & $0.0208097$ & $0.0171735$ & $%
0.0172249$ & $0.0171740$ \\ 
& $0.050$ & $0.0117234$ & $0.0119292$ & $0.0117365$ & $0.0088961$ & $%
0.0091019$ & $0.0089134$ \\ 
& $0.075$ & $0.0050143$ & $0.0054773$ & $0.0050945$ & $0.0030849$ & $%
0.0035478$ & $0.0031884$ \\ 
$4d$ & $0.025$ & $0.0203012$ & $0.0204555$ & $0.0203017$ & $0.0182106$ & $%
0.0183649$ & $0.0182115$ \\ 
& $0.050$ & $0.0109569$ & $0.0115742$ & $0.0109904$ & $0.0094775$ & $%
0.0100947$ & $0.0095167$ \\ 
& $0.075$ & $0.0038158$ & $0.0052047$ & $0.0040331$ & $0.0028919$ & $%
0.0042808$ & $0.0031399$ \\ 
$4f$ & $0.025$ & $0.0199801$ & $0.0202887$ & $0.0199797$ & $0.0186136$ & $%
0.0189223$ & $0.0186137$ \\ 
& $0.050$ & $0.0101938$ & $0.0114284$ & $0.0102393$ & $0.0093507$ & $%
0.0105852$ & $0.0094015$ \\ 
& $0.075$ & $0.0023157$ & $0.0050935$ & $0.0026443$ & $0.0018749$ & $%
0.0046527$ & $0.0022307$ \\ 
$5p$ & $0.025$ & $0.0098062$ & $0.0098576$ & $0.0098079$ & $0.0080793$ & $%
0.0081308$ & $0.0080816$ \\ 
$5d$ & $0.025$ & $0.0095094$ & $0.0096637$ & $0.0095141$ & $0.0085359$ & $%
0.0086902$ & $0.0085415$ \\ 
$5f$ & $0.025$ & $0.0092751$ & $0.0095837$ & $0.0092825$ & $0.0086536$ & $%
0.0089622$ & $0.0086619$ \\ 
$5g$ & $0.025$ & $0.0090254$ & $0.0095398$ & $0.0090330$ & $0.0086066$ & $%
0.0091210$ & $0.0086150$ \\ 
$6p$ & $0.025$ & $0.0043537$ & $0.0044051$ & $0.0043583$ & $0.0034820$ & $%
0.0035334$ & $0.0034876$ \\ 
$6d$ & $0.025$ & $0.0041518$ & $0.0043061$ & $0.0041650$ & $0.0036666$ & $%
0.0038209$ & $0.0036813$ \\ 
$6f$ & $0.025$ & $0.0039566$ & $0.0042652$ & $0.0039803$ & $0.0036520$ & $%
0.0039606$ & $0.0036774$ \\ 
$6g$ & $0.025$ & $0.0037284$ & $0.0042428$ & $0.0037611$ & $0.0035278$ & $%
0.0040422$ & $0.0035623$%
\end{tabular}%
\end{table}

\begin{table}[tbp]
\caption{The ro-vibrational energy spectra ($-E_{nl}$) (in $eV$) for $%
\mathrm{HCl}$ and $\mathrm{CH}$ for $2p,3p,3d,4p,4d,4f,5p,5d,5f,5g,6p,6d,6f$%
and $6g$ states with $\hbar c=1973.29$ $eV$ $A^{\circ },$ $\protect\mu _{%
\mathrm{HCl}}=0.9801045$ $amu,$ $\protect\mu _{\mathrm{CH}}=0.929931$ $amu$
and $A=2b.$}%
\begin{tabular}{llllllll}
\tableline states & $1/b$\tablenotetext[1]{$b$ is in $pm$.}\tablenotemark[1]
& HCl$/$ $\alpha =0,1$ & $\alpha =0.75$ & $\alpha =1.5$ & CH$/$ $\alpha =0,1$
& $\alpha =0.75$ & $\alpha =1.5$ \\ 
\tableline$2p$ & $0.025$ & $4.80933$ & $5.14059$ & $3.83734$ & $5.06882$ & $%
5.41795$ & $4.04438$ \\ 
& $0.050$ & $4.30960$ & $4.61553$ & $3.41382$ & $4.54212$ & $4.86455$ & $%
3.59801$ \\ 
& $0.075$ & $3.83214$ & $4.11362$ & $3.00987$ & $4.03890$ & $4.33556$ & $%
3.17226$ \\ 
& $0.100$ & $3.37695$ & $3.63486$ & $2.42549$ & $3.55915$ & $3.83097$ & $%
2.76714$ \\ 
$3p$ & $0.025$ & $1.86414$ & $1.95674$ & $1.57439$ & $1.96472$ & $2.06231$ & 
$1.65934$ \\ 
& $0.050$ & $1.41439$ & $1.49539$ & $1.16292$ & $1.49071$ & $1.57608$ & $%
1.22566$ \\ 
& $0.075$ & $1.02023$ & $1.08964$ & $0.80704$ & $1.07528$ & $1.14843$ & $%
0.85058$ \\ 
& $0.100$ & $0.68166$ & $0.73947$ & $0.50674$ & $0.71844$ & $0.77937$ & $%
0.53409$ \\ 
$3d$ & $0.025$ & $1.85975$ & $1.90971$ & $1.68385$ & $1.96010$ & $2.01275$ & 
$1.77470$ \\ 
& $0.050$ & $1.39684$ & $1.43694$ & $1.25588$ & $1.47221$ & $1.51447$ & $%
1.32363$ \\ 
& $0.075$ & $0.98074$ & $1.01201$ & $0.87106$ & $1.03366$ & $1.06661$ & $%
0.91806$ \\ 
& $0.100$ & $0.61146$ & $0.63492$ & $0.52941$ & $0.64445$ & $0.66917$ & $%
0.55798$ \\ 
$4p$ & $0.025$ & $0.85082$ & $0.88753$ & $0.73246$ & $0.89672$ & $0.93542$ & 
$0.77198$ \\ 
& $0.050$ & $0.47104$ & $0.50001$ & $0.37942$ & $0.496459$ & $0.526989$ & $%
0.399896$ \\ 
& $0.075$ & $0.19351$ & $0.21387$ & $0.13157$ & $0.203948$ & $0.225404$ & $%
0.138671$ \\ 
$4d$ & $0.025$ & $0.84643$ & $0.86586$ & $0.77669$ & $0.892099$ & $0.912577$
& $0.818599$ \\ 
& $0.050$ & $0.45349$ & $0.46732$ & $0.40422$ & $0.477960$ & $0.492531$ & $%
0.426029$ \\ 
& $0.075$ & $0.15402$ & $0.16275$ & $0.12334$ & $0.162325$ & $0.171527$ & $%
0.129997$ \\ 
$4f$ & $0.025$ & $0.83985$ & $0.85216$ & $0.79388$ & $0.885162$ & $0.898138$
& $0.836716$ \\ 
& $0.050$ & $0.42716$ & $0.43477$ & $0.39881$ & $0.450211$ & $0.458228$ & $%
0.420329$ \\ 
& $0.075$ & $0.094777$ & $0.098765$ & $0.079967$ & $0.099891$ & $0.104094$ & 
$0.084281$ \\ 
$5p$ & $0.025$ & $0.40099$ & $0.41824$ & $0.34459$ & $0.422623$ & $0.440805$
& $0.363181$ \\ 
$5d$ & $0.025$ & $0.39660$ & $0.40558$ & $0.36406$ & $0.417998$ & $0.427463$
& $0.383705$ \\ 
$5f$ & $0.025$ & $0.390018$ & $0.395586$ & $0.36908$ & $0.411061$ & $%
0.416929 $ & $0.388993$ \\ 
$5g$ & $0.025$ & $0.381242$ & $0.38494$ & $0.367077$ & $0.401811$ & $%
0.405709 $ & $0.386882$ \\ 
$6p$ & $0.025$ & $0.176998$ & $0.18569$ & $0.14851$ & $0.186548$ & $0.195706$
& $0.156521$ \\ 
$6d$ & $0.025$ & $0.172610$ & $0.17708$ & $0.15638$ & $0.181923$ & $0.186631$
& $0.164820$ \\ 
$6f$ & $0.025$ & $0.166028$ & $0.168752$ & $0.155759$ & $0.174986$ & $%
0.177856$ & $0.164163$ \\ 
$6g$ & $0.025$ & $0.157252$ & $0.15902$ & $0.150462$ & $0.165736$ & $%
0.167600 $ & $0.158580$%
\end{tabular}%
\end{table}

\begin{table}[tbp]
\caption{The ro-vibrational energy spectra $(-E_{nl})$ (in $eV$) for $%
\mathrm{LiH}$ and $\mathrm{CO}$ for $2p,3p,3d,4p,4d,4f,5p,5d,5f,5g,6p,6d,6f$%
and $6g$ states with $\protect\mu _{\mathrm{LiH}}=0.8801221$ $amu,$ $\protect%
\mu _{\mathrm{CO}}=6.8606719$ $amu$ and $A=2b.$}%
\begin{tabular}{llllllll}
\tableline states & $1/b$\tablenotemark[1]\tablenotetext[1]{$b$ is in $pm$.}
& LiH$/$ $\alpha =0,1$ & $\alpha =0.75$ & $\alpha =1.5$ & CO$/$ $\alpha =0,1$
& $\alpha =0.75$ & $\alpha =1.5$ \\ 
\tableline$2p$ & $0.025$ & $5.35568$ & $5.72457$ & $4.27326$ & $0.687053$ & $%
0.734377$ & $0.548196$ \\ 
& $0.050$ & $4.79918$ & $5.13985$ & $3.80163$ & $0.615663$ & $0.659367$ & $%
0.487693$ \\ 
& $0.075$ & $4.26747$ & $4.58092$ & $3.35179$ & $0.547453$ & $0.587664$ & $%
0.429985$ \\ 
& $0.100$ & $3.76057$ & $4.04778$ & $2.92374$ & $0.482425$ & $0.519270$ & $%
0.375073$ \\ 
$3p$ & $0.025$ & $2.07591$ & $2.17902$ & $1.75324$ & $0.266308$ & $0.279536$
& $0.224915$ \\ 
& $0.050$ & $1.57507$ & $1.66527$ & $1.29503$ & $0.202058$ & $0.213629$ & $%
0.166133$ \\ 
& $0.075$ & $1.13613$ & $1.21342$ & $0.89872$ & $0.145749$ & $0.155664$ & $%
0.115292$ \\ 
& $0.100$ & $0.759101$ & $0.823478$ & $0.564311$ & $0.097381$ & $0.105640$ & 
$0.072393$ \\ 
$3d$ & $0.025$ & $2.07102$ & $2.12665$ & $1.87514$ & $0.265681$ & $0.272818$
& $0.240553$ \\ 
& $0.050$ & $1.55552$ & $1.60018$ & $1.39854$ & $0.199550$ & $0.205279$ & $%
0.179412$ \\ 
& $0.075$ & $1.09215$ & $1.12698$ & $0.970015$ & $0.140107$ & $0.144574$ & $%
0.124439$ \\ 
& $0.100$ & $0.680918$ & $0.707045$ & $0.589556$ & $0.087352$ & $0.090703$ & 
$0.075631$ \\ 
$4p$ & $0.025$ & $0.947473$ & $0.988358$ & $0.815668$ & $0.121547$ & $%
0.126792$ & $0.104638$ \\ 
& $0.050$ & $0.524555$ & $0.556813$ & $0.422528$ & $0.067293$ & $0.071431$ & 
$0.054204$ \\ 
& $0.075$ & $0.215490$ & $0.238160$ & $0.146518$ & $0.027644$ & $0.030552$ & 
$0.018796$ \\ 
$4d$ & $0.025$ & $0.942586$ & $0.964223$ & $0.864926$ & $0.120920$ & $%
0.123695$ & $0.110957$ \\ 
& $0.050$ & $0.505009$ & $0.520405$ & $0.450139$ & $0.064785$ & $0.066760$ & 
$0.057746$ \\ 
& $0.075$ & $0.171512$ & $0.181234$ & $0.137354$ & $0.022002$ & $0.023250$ & 
$0.017620$ \\ 
$4f$ & $0.025$ & $0.935256$ & $0.948967$ & $0.884069$ & $0.119979$ & $%
0.121738$ & $0.113413$ \\ 
& $0.050$ & $0.475690$ & $0.484161$ & $0.444117$ & $0.061024$ & $0.062111$ & 
$0.056974$ \\ 
& $0.075$ & $0.105544$ & $0.109984$ & $0.089051$ & $0.013540$ & $0.014109$ & 
$0.011424$ \\ 
$5p$ & $0.025$ & $0.446540$ & $0.465751$ & $0.383735$ & $0.057284$ & $%
0.059749$ & $0.049227$ \\ 
$5d$ & $0.025$ & $0.441654$ & $0.451655$ & $0.405420$ & $0.056658$ & $%
0.057941$ & $0.052009$ \\ 
$5f$ & $0.025$ & $0.434324$ & $0.440525$ & $0.411008$ & $0.055717$ & $%
0.056513$ & $0.052726$ \\ 
$5g$ & $0.025$ & $0.424551$ & $0.428669$ & $0.408777$ & $0.054464$ & $%
0.054992$ & $0.052440$ \\ 
$6p$ & $0.025$ & $0.197105$ & $0.206782$ & $0.165379$ & $0.025286$ & $%
0.026527$ & $0.021216$ \\ 
$6d$ & $0.025$ & $0.192219$ & $0.197193$ & $0.174148$ & $0.024659$ & $%
0.025297$ & $0.022341$ \\ 
$6f$ & $0.025$ & $0.184889$ & $0.187922$ & $0.173454$ & $0.023718$ & $%
0.024108$ & $0.022252$ \\ 
$6g$ & $0.025$ & $0.175116$ & $0.177085$ & $0.167554$ & $0.022465$ & $%
0.022717$ & $0.021495$%
\end{tabular}%
\end{table}
\begin{table}[tbp]
\caption{The ro-vibrational energy spectra $(-E_{nl})$ (in $eV$) for $%
\mathrm{NO}$ and $\mathrm{O}_{2}$ for $%
2p,3p,3d,4p,4d,4f,5p,5d,5f,5g,6p,6d,6f $and $6g$ states with $\protect\mu _{%
\mathrm{NO}}=7.468441$ $amu,$ $\protect\mu _{\mathrm{O}_{2}}=7.997457504$ $%
amu$ and $A=2b.$}%
\begin{tabular}{llllllll}
\tableline states & $1/b$\tablenotemark[1]\tablenotetext[1]{$b$ is in $pm$.}
& NO$/$ $\alpha =0,1$ & $\alpha =0.75$ & $\alpha =1.5$ & O$_{2}/$ $\alpha
=0,1$ & $\alpha =0.75$ & $\alpha =1.5$ \\ 
\tableline$2p$ & $0.025$ & $0.631142$ & 0$.674615$ & $0.503585$ & $0.589393$
& $0.629990$ & $0.470274$ \\ 
& $0.050$ & $0.565561$ & $0.605709$ & $0.448005$ & $0.528150$ & $0.565642$ & 
$0.418370$ \\ 
& $0.075$ & $0.502903$ & $0.539841$ & $0.394993$ & $0.469637$ & $0.504132$ & 
$0.368865$ \\ 
& $0.100$ & $0.443166$ & $0.477013$ & $0.344550$ & $0.413852$ & $0.445459$ & 
$0.321759$ \\ 
$3p$ & $0.025$ & $0.244637$ & $0.256788$ & $0.206612$ & $0.228454$ & $%
0.239802$ & $0.192945$ \\ 
& $0.050$ & $0.185615$ & $0.196245$ & $0.152613$ & $0.173337$ & $0.183263$ & 
$0.142518$ \\ 
& $0.075$ & $0.133888$ & $0.142996$ & $0.105910$ & $0.125032$ & $0.133537$ & 
$0.098904$ \\ 
& $0.100$ & $0.089457$ & $0.097043$ & $0.066502$ & $0.083539$ & $0.090624$ & 
$0.062103$ \\ 
$3d$ & $0.025$ & $0.244061$ & $0.250617$ & $0.220977$ & $0.227917$ & $%
0.234039$ & $0.206360$ \\ 
& $0.050$ & $0.183311$ & $0.188574$ & $0.164812$ & $0.171186$ & $0.176100$ & 
$0.153910$ \\ 
& $0.075$ & $0.128706$ & $0.132809$ & $0.114312$ & $0.120192$ & $0.124024$ & 
$0.106750$ \\ 
& $0.100$ & $0.080243$ & $0.083322$ & $0.069477$ & $0.074935$ & $0.077810$ & 
$0.064881$ \\ 
$4p$ & $0.025$ & $0.111655$ & $0.116474$ & $0.096123$ & $0.104270$ & $%
0.108769$ & $0.089764$ \\ 
& $0.050$ & $0.061816$ & $0.065618$ & $0.049793$ & $0.057727$ & $0.061277$ & 
$0.046499$ \\ 
& $0.075$ & $0.025395$ & $0.028066$ & $0.017267$ & $0.023715$ & $0.026210$ & 
$0.016124$ \\ 
$4d$ & $0.025$ & $0.111080$ & $0.113629$ & $0.101928$ & $0.103732$ & $%
0.106113$ & $0.095185$ \\ 
& $0.050$ & $0.059513$ & $0.061327$ & $0.053047$ & $0.055576$ & $0.057271$ & 
$0.049538$ \\ 
& $0.075$ & $0.020212$ & $0.021358$ & $0.016187$ & $0.018875$ & $0.019945$ & 
$0.015116$ \\ 
$4f$ & $0.025$ & $0.110216$ & $0.111831$ & $0.104184$ & $0.102925$ & $%
0.104434$ & $0.097292$ \\ 
& $0.050$ & $0.056058$ & $0.057056$ & $0.052337$ & $0.052350$ & $0.053282$ & 
$0.048875$ \\ 
& $0.075$ & $0.012438$ & $0.012961$ & $0.010494$ & $0.011615$ & $0.012104$ & 
$0.009800$ \\ 
$5p$ & $0.025$ & $0.052623$ & $0.054887$ & $0.045221$ & $0.049142$ & $%
0.051256$ & $0.042230$ \\ 
$5d$ & $0.025$ & $0.052047$ & $0.053225$ & $0.047777$ & $0.048604$ & $%
0.049705$ & $0.044617$ \\ 
$5f$ & $0.025$ & $0.051183$ & $0.051914$ & $0.048435$ & $0.047797$ & $%
0.048480$ & $0.045231$ \\ 
$5g$ & $0.025$ & $0.050031$ & $0.050517$ & $0.048173$ & $0.046722$ & $%
0.047175$ & $0.044986$ \\ 
$6p$ & $0.025$ & $0.023228$ & $0.024368$ & $0.019489$ & $0.021691$ & $%
0.022756$ & $0.018200$ \\ 
$6d$ & $0.025$ & $0.022652$ & $0.023238$ & $0.020523$ & $0.021154$ & $%
0.021701$ & $0.019165$ \\ 
$6f$ & $0.025$ & $0.021788$ & $0.022146$ & $0.020441$ & $0.020347$ & $%
0.020681$ & $0.019089$ \\ 
$6g$ & $0.025$ & $0.020637$ & $0.020869$ & $0.019746$ & $0.019272$ & $%
0.019488$ & $0.018439$%
\end{tabular}%
\end{table}
\begin{table}[tbp]
\caption{The ro-vibrational energy spectra $(-E_{nl})$ (in $eV$) for $%
\mathrm{I}_{2}$ and $\mathrm{N}_{2}$ for $%
2p,3p,3d,4p,4d,4f,5p,5d,5f,5g,6p,6d,6f$and $6g$ states with $\protect\mu _{%
\mathrm{I}_{2}}=63.45223502$ $amu,$ $\protect\mu _{\mathrm{N}_{2}}=7.00335$ $%
amu$ and $A=2b.$}%
\begin{tabular}{llllllll}
\tableline states & $1/b$\tablenotemark[1]\tablenotetext[1]{$b$ is in $pm$.}
& I$_{2}/$ $\alpha =0,1$ & $\alpha =0.75$ & $\alpha =1.5$ & N$_{2}/$ $\alpha
=0,1$ & $\alpha =0.75$ & $\alpha =1.5$ \\ 
\tableline$2p$ & $0.025$ & $0.0742866$ & 0$.0794033$ & $0.0592729$ & $%
0.673056$ & $0.719416$ & $0.537028$ \\ 
& $0.050$ & $0.0665676$ & $0.0712930$ & $0.0527310$ & $0.603120$ & $0.645934$
& $0.477757$ \\ 
& $0.075$ & $0.0591925$ & $0.0635403$ & $0.0464914$ & $0.536300$ & $0.575692$
& $0.421225$ \\ 
& $0.100$ & $0.0521615$ & $0.0561452$ & $0.0405541$ & $0.472597$ & $0.508691$
& $0.367431$ \\ 
$3p$ & $0.025$ & $0.0287942$ & $0.0302244$ & $0.0243186$ & $0.260883$ & $%
0.273841$ & $0.220333$ \\ 
& $0.050$ & $0.0218472$ & $0.0230983$ & $0.0179628$ & $0.197941$ & $0.209277$
& $0.162748$ \\ 
& $0.075$ & $0.0157589$ & $0.0168309$ & $0.0124658$ & $0.142780$ & $0.152493$
& $0.112943$ \\ 
& $0.100$ & $0.0105292$ & $0.0114221$ & $0.0078274$ & $0.095397$ & $0.103488$
& $0.070918$ \\ 
$3d$ & $0.025$ & $0.0287264$ & $0.0294980$ & $0.0260094$ & $0.260269$ & $%
0.267260$ & $0.235652$ \\ 
& $0.050$ & $0.0215761$ & $0.0221955$ & $0.0193987$ & $0.195485$ & $0.201097$
& $0.175757$ \\ 
& $0.075$ & $0.0151489$ & $0.0156319$ & $0.0134547$ & $0.137253$ & $0.141629$
& $0.121903$ \\ 
& $0.100$ & $0.0094448$ & $0.0098072$ & $0.0081775$ & $0.085572$ & $0.088855$
& $0.074090$ \\ 
$4p$ & $0.025$ & $0.0131420$ & $0.0137091$ & $0.0113138$ & $0.119070$ & $%
0.124209$ & $0.102506$ \\ 
& $0.050$ & $0.0072759$ & $0.0072330$ & $0.0058607$ & $0.065922$ & $0.069976$
& $0.053100$ \\ 
& $0.075$ & $0.0029890$ & $0.0033034$ & $0.0020323$ & $0.027081$ & $0.029930$
& $0.018413$ \\ 
$4d$ & $0.025$ & $0.0130743$ & $0.0133744$ & $0.0119971$ & $0.118456$ & $%
0.121175$ & $0.108697$ \\ 
& $0.050$ & $0.0070048$ & $0.0072183$ & $0.0062437$ & $0.063465$ & $0.065400$
& $0.056570$ \\ 
& $0.075$ & $0.0023790$ & $0.0025138$ & $0.0019052$ & $0.021554$ & $0.022776$
& $0.017261$ \\ 
$4f$ & $0.025$ & $0.0129726$ & $0.0131628$ & $0.0122626$ & $0.117535$ & $%
0.119258$ & $0.111102$ \\ 
& $0.050$ & $0.0065981$ & $0.0067156$ & $0.0061602$ & $0.059781$ & $0.060845$
& $0.055813$ \\ 
& $0.075$ & $0.0014640$ & $0.0015256$ & $0.0012352$ & $0.013264$ & $0.013822$
& $0.011191$ \\ 
$5p$ & $0.025$ & $0.0061938$ & $0.0064603$ & $0.0053226$ & $0.056117$ & $%
0.058532$ & $0.048225$ \\ 
$5d$ & $0.025$ & $0.0061260$ & $0.0062647$ & $0.0056234$ & $0.055503$ & $%
0.056760$ & $0.050950$ \\ 
$5f$ & $0.025$ & $0.0060243$ & $0.0061104$ & $0.0057009$ & $0.054582$ & $%
0.055361$ & $0.051652$ \\ 
$5g$ & $0.025$ & $0.0058888$ & $0.0059459$ & $0.0056700$ & $0.053354$ & $%
0.053872$ & $0.051372$ \\ 
$6p$ & $0.025$ & $0.0027340$ & $0.0028682$ & $0.0022939$ & $0.024771$ & $%
0.025987$ & $0.020783$ \\ 
$6d$ & $0.025$ & $0.0026662$ & $0.0027352$ & $0.0024155$ & $0.024156$ & $%
0.024782$ & $0.021885$ \\ 
$6f$ & $0.025$ & $0.0025645$ & $0.0026066$ & $0.0024059$ & $0.023235$ & $%
0.023616$ & $0.021798$ \\ 
$6g$ & $0.025$ & $0.0024290$ & $0.0024563$ & $0.0023241$ & $0.022007$ & $%
0.022255$ & $0.021057$%
\end{tabular}%
\end{table}
\begin{table}[tbp]
\caption{The ro-vibrational energy spectra $(-E_{nl})$ (in $eV$) for $%
\mathrm{H}_{2}$ and $\mathrm{Ar}_{2}$ for $%
2p,3p,3d,4p,4d,4f,5p,5d,5f,5g,6p,6d,6f$and $6g$ states with $\protect\mu _{%
\mathrm{H}_{2}}=0.50407$ $amu,$ $\protect\mu _{\mathrm{Ar}_{2}}=19.9812$ $%
amu $ and $A=2b.$}%
\begin{tabular}{llllllll}
\tableline states & $1/b$\tablenotemark[1]\tablenotetext[1]{$b$ is in $pm$.}
& H$_{2}/$ $\alpha =0,1$ & $\alpha =0.75$ & $\alpha =1.5$ & Ar$_{2}/$ $%
\alpha =0,1$ & $\alpha =0.75$ & $\alpha =1.5$ \\ 
\tableline$2p$ & $0.025$ & $9.35118$ & 9$.99528$ & $7.46126$ & $0.235904$ & $%
0.252153$ & $0.188227$ \\ 
& $0.050$ & $8.37951$ & $8.97435$ & $6.63777$ & $0.211392$ & $0.226398$ & $%
0.167452$ \\ 
& $0.075$ & $7.45114$ & $7.99844$ & $5.85233$ & $0.187972$ & $0.201778$ & $%
0.147638$ \\ 
& $0.100$ & $6.56608$ & $7.06755$ & $5.10495$ & $0.165644$ & $0.178295$ & $%
0.128784$ \\ 
$3p$ & $0.025$ & $3.62460$ & $3.80464$ & $3.06122$ & $0.091439$ & $0.095981$
& $0.077226$ \\ 
& $0.050$ & $2.75012$ & $2.90761$ & $2.26116$ & $0.069378$ & $0.073351$ & $%
0.057043$ \\ 
& $0.075$ & $1.98372$ & $2.11867$ & $1.56919$ & $0.050044$ & $0.053448$ & $%
0.039586$ \\ 
& $0.100$ & $1.32541$ & $1.43782$ & $0.98531$ & $0.033437$ & $0.036272$ & $%
0.024857$ \\ 
$3d$ & $0.025$ & $3.61607$ & $3.71320$ & $3.27405$ & $0.0912234$ & $%
0.0936738 $ & $0.0825953$ \\ 
& $0.050$ & $2.71599$ & $2.79396$ & $2.44190$ & $0.0685169$ & $0.0704839$ & $%
0.0616024$ \\ 
& $0.075$ & $1.90694$ & $1.96773$ & $1.69368$ & $0.0481067$ & $0.0496405$ & $%
0.0427268$ \\ 
& $0.100$ & $1.18890$ & $1.23452$ & $102938$ & $0.0299927$ & $0.0311436$ & $%
0.0259685$ \\ 
$4p$ & $0.025$ & $1.65432$ & $1.72570$ & $1.42418$ & $0.041734$ & $0.043535$
& $0.035928$ \\ 
& $0.050$ & $0.91589$ & $0.97221$ & $0.73775$ & $0.023105$ & $0.024526$ & $%
0.018611$ \\ 
& $0.075$ & $0.37625$ & $0.41584$ & $0.25583$ & $0.0094918$ & $0.0104904$ & $%
0.0064538$ \\ 
$4d$ & $0.025$ & $1.64578$ & $1.68356$ & $1.51019$ & $0.0415186$ & $%
0.0424716 $ & $0.0380978$ \\ 
& $0.050$ & $0.88176$ & $0.90864$ & $0.78596$ & $0.0222444$ & $0.0229225$ & $%
0.0198275$ \\ 
& $0.075$ & $0.29946$ & $0.31644$ & $0.23982$ & $0.0075547$ & $0.0079829$ & $%
0.0060501$ \\ 
$4f$ & $0.025$ & $1.63299$ & $1.65693$ & $1.54361$ & $0.0411957$ & $%
0.0417996 $ & $0.0389410$ \\ 
& $0.050$ & $0.83057$ & $0.84536$ & $0.77544$ & $0.0209530$ & $0.0213261$ & $%
0.0195623$ \\ 
& $0.075$ & $0.18428$ & $0.19204$ & $0.15549$ & $0.0046490$ & $0.0048445$ & $%
0.0039225$ \\ 
$5p$ & $0.025$ & $0.77967$ & $0.81322$ & $0.67001$ & $0.0196690$ & $%
0.0205152 $ & $0.0169026$ \\ 
$5d$ & $0.025$ & $0.77114$ & $0.78860$ & $0.70788$ & $0.0194538$ & $%
0.0198943 $ & $0.0178578$ \\ 
$5f$ & $0.025$ & $0.75834$ & $0.76917$ & $0.71763$ & $0.0191309$ & $%
0.0194040 $ & $0.0181039$ \\ 
$5g$ & $0.025$ & $0.74128$ & $0.74847$ & $0.71374$ & $0.0187004$ & $%
0.0188818 $ & $0.0180056$ \\ 
$6p$ & $0.025$ & $0.34415$ & $0.36105$ & $0.28876$ & $0.0086820$ & $%
0.0091082 $ & $0.0072845$ \\ 
$6d$ & $0.025$ & $0.33562$ & $0.34430$ & $0.30407$ & $0.0084667$ & $%
0.0086859 $ & $0.0076708$ \\ 
$6f$ & $0.025$ & $0.32282$ & $0.32812$ & $0.30286$ & $0.0081439$ & $%
0.0827750 $ & $0.0076402$ \\ 
$6g$ & $0.025$ & $0.30576$ & $0.30920$ & $0.29256$ & $0.0077134$ & $%
0.0078001 $ & $0.0073804$%
\end{tabular}%
\end{table}


\begin{thebibliography}{99}
\bibitem{1} L.~I.~Schiff, Quantum Mechanics 3rd edn. (McGraw-Hill Book Co.,
New York, 1968).

\bibitem{2} L.~D.~Landau and E.~M.~Lifshitz, Quantum Mechanics,
Non-relativistic Theory, 3rd edn. (Pergamon, New York, 1977).

\bibitem{3} M.~M.~Neito, Am. J. Phys. 47 (1979) 1067.

\bibitem{4} S. Ikhdair and R. Sever, J. Mol. Struct.-Theochem 806 (2007) 155.

\bibitem{5} S.M. Ikhdair and R. Sever, J. Mol. Struct.-Theochem 855 (2008)
13.

\bibitem{6} C.L. Pekeris, Phys. Rev. 45 (1934) 98; C. Berkdemir, Nucl. Phys.
A 770 (2006) 32.

\bibitem{7} W.-C. Qiang and S.-H. Dong, Phys. Lett. A 363 (2007) 169.

\bibitem{8} C. Berkdemir and J. Han, Chem. Phys. Lett. 409 (2005) 203; C.
Berkdemir, A. Berkdemir and J. Han, Chem. Phys. Lett. 417 (2006) 326.

\bibitem{9} R.L. Greene and C. Aldrich, Phys. Rev. A 14 (1976) 2363; B. G%
\"{o}n\"{u}l and \.{I}. Zorba, Phys. Lett. A 269 (2000) 83.

\bibitem{10} P.M. Morse, Phys. Rev. 34 (1929) 57; S.M. Ikhdair, Chem. Phys.
361 (1-2) (2009) 9.

\bibitem{11} L. Hulth\'{e}n, Ark. Mat. Astron. Fys. A 28 (1942) 5; S.M.
Ikhdair and R. Sever, J. Math. Chem. 42 (3) (2007) 461.

\bibitem{12} \"{O}. Ye\c{s}ilta\c{s}, Phys. Scr. 75 (2007) 41.

\bibitem{13} S.M. Ikhdair and R. Sever, Int. J. Mod. Phys. A 25 (2010) 3941;
S.M. Ikhdair and R. Sever, Cent. Eur. J. Phys. 8 (2010) 652; S.M. Ikhdair
and R. Sever, Int. J. Theor. Phys. 46 (2007) 1643.

\bibitem{14} S.M. Ikhdair, Chin. J. Phys. 46 (2008) 291; S.M. Ikhdair and R.
Sever, Int. J. Mod. Phys. C 18 (2007) 1571; S.M. Ikhdair and R. Sever, Int.
J. Mod. Phys. C 19 (2008) 221; S.M. Ikhdair and R. Sever, Cent. Eur. J.
Phys. 5 (2007) 516; S.M. Ikhdair and R. Sever, Cent. Eur. J. Phys. 6 (2008)
141, 685; 697; S.M. Ikhdair and R. Sever, Int. J. Mod. Phys. C 19 (2008)
1425; S.M. Ikhdair and R. Sever, J. Math. Chem. 45 (2009) 1137.

\bibitem{15} S.M. Ikhdair, J. Math. Phys. 51 (2010) 023525; N. Rosen and
P.M. Morse, Phys. Rev. 42 (1932) 210.

\bibitem{16} M.F. Manning, Phys. Rev. 44 (1933) 951; M.F. Manning and N.
Rosen, Phys. Rev. 44 (1933) 953.

\bibitem{17} A. Diaf, A. Chouchaoui and R.L. Lombard, Ann. Phys. (Paris) 317
(2005) 354.

\bibitem{18} S.-H. Dong and J. Garcia-Ravelo, Phys. Scr. 75 (2007) 307.

\bibitem{19} W.-C. Qiang and S. H. Dong, Phys. Lett. A 368 (2007) 13.

\bibitem{20} C.-S. Jia \textit{et al}., J. Phys. A: Math. Gen. 37 (2004)
11275; \ C.-S. Jia \textit{et al}., Phys. Lett. A 311 (2003) 115.

\bibitem{21} H. E\u{g}rifes, D. Demirhan and F. B\"{u}y\"{u}kk\i l\i \c{c},
Phys. Lett. A 275 (2000) 229.

\bibitem{22} W.-C. Qiang, Chin. Phys. 12 (2003) 1054; \textit{ibid.} 13
(2004) 575; L.-Z. Yi, Y.-F. Diao, J.-Y. Liu and C.-S. Jia, Phys. Lett. A 333
(2004) 212; G.-F. Wei and S.-H. Dong, Phys. Lett. A 373 (2008) 49.

\bibitem{23} R.J. Le Roy and R.B. Bernstein, J. Chem. Phys. 52 (1970) 3869.

\bibitem{24} J. Cai, P. Cai and A. Inomata, Phys. Rev. A 34 (1986) 4621.

\bibitem{25} G.F. Wei, C.Y. Long and S.H. Dong, Phys. Lett. A 372 (2008)
2592.

\bibitem{26} S.M. Ikhdair and R. Sever, Ann. Phys. (Berlin) 17 (2008) 897;
S.M. Ikhdair and R. Sever, Phys. Scr. 79 (2009) 035002.

\bibitem{27} S.M. Ikhdair, Eur. Phys. J. A 39 (2009) 307.

\bibitem{28} W. Lucha and F.F. Sch\"{o}berl, Int. J. Mod. Phys. C 10 (1999)
607.

\bibitem{29} S.M. Ikhdair and R. Sever, to appear in the Int. J. Mod. Phys.
B (2011) [arXiv:0909.0623]; S.M. Ikhdair and R. Sever, Ann. Phys. (Berlin)
18 (2009) 747.

\bibitem{30} A.F. Nikiforov and V.B. Uvarov, Special Functions of
Mathematical Physics (Birkhauser, Bassel, 1988).

\bibitem{31} S.M. Ikhdair and R. Sever, Appl. Math. Comp. 216 (2010) 911.

\bibitem{32} G.-F. Wei and S.-H. Dong, Phys. Lett. A 373 (2008) 49.

\bibitem{33} S.M. Ikhdair, Chem. Phys. 361 (2009) 9.

\bibitem{34} F. Benamira, L. Guechi and A. Zouache, Phys. Scr. 80 (2009)
017001.

\bibitem{35} I.S. Gradshtein and I.M. Ryzhik, Tables and integrals, series
and products (New York, Academic, 1969).
\end{thebibliography}
\end{document}